\documentclass[aps,prl,twocolumn, amsmath,superscriptaddress,nofootinbib,amssymb]{revtex4}
\usepackage{graphicx}
\usepackage[colorlinks, linkcolor=blue]{hyperref}
\usepackage{verbatim}
\usepackage{latexsym}
\usepackage{amsmath}
\usepackage{amssymb}
\usepackage{hyperref}

\begin{document}

\title{Hydrodynamic flows of non-Fermi liquids: magnetotransport and bilayer drag}

\author{Aavishkar A. Patel}
\affiliation{Department of Physics, Harvard University, Cambridge MA 02138, USA}

\author{Richard A. Davison}
\affiliation{Department of Physics, Harvard University, Cambridge MA 02138, USA}

\author{Alex Levchenko}
\affiliation{Department of Physics, University of Wisconsin-Madison, Madison, Wisconsin 53706, USA}

\begin{abstract}
We consider a hydrodynamic description of transport for generic two dimensional electron systems that lack Galilean invariance and do not fall into the category of Fermi liquids. We study magnetoresistance and show that it is governed only by the electronic viscosity provided that the wavelength of the underlying disorder potential is large compared to the microscopic equilibration length. We also derive the Coulomb drag transresistance for double-layer non-Fermi liquid systems in the hydrodynamic regime. As an example, we consider frictional drag between two quantum Hall states with half-filled lowest Landau levels, each described by a Fermi surface of composite fermions coupled to a $U(1)$ gauge field. We contrast our results to prior calculations of drag of Chern-Simons composite particles and place our findings in the context of available experimental data.
\end{abstract}

\maketitle

\textit{Introduction}. Hydrodynamic flow of electrons can occur in solid state systems provided that the microscopic length scale of momentum-conserving electron-electron collisions is sufficiently short \cite{Gurzhi:1968}. Under this condition the electron liquid attains local equilibrium and can be described in terms of slow variables associated with conserved quantities such as momentum and energy.  However, this transport regime was hard to realize experimentally as typically electron-impurity scattering degrades electron momentum, whereas electron-phonon collisions violate both momentum and energy conservations of the electron liquid. Early evidence for the so-called hydrodynamic Gurzhi effect, related to the negative temperature derivative of resistivity, was reported in thin potassium wires \cite{Yu:1984}, and later in the electrostatically defined wires in the two dimensional electron gas of (Al,Ga)As heterostructures \cite{Molenkamp:1995}. The recent surge of experiments devoted to revealing hydrodynamic regimes of electronic transport is mainly focused on measurements conducted on graphene \cite{Bandurin:2016,Crossno:2016}.  
 
In the context of transport theories, a hydrodynamic description is powerful as it accurately describes most liquids. All microscopic details of the system at hand are then encoded into a handful of kinetic coefficients such as viscosities and thermal conductivity. In certain cases the latter can be controllably derived from the linearized Boltzmann kinetic equation by following the perturbative Chapman-Enskog procedure developed originally for gases. However, we have examples now where this kind of microscopic approach has to be substantially revisited. Deriving hydrodynamics for linearly dispersing electronic excitations in graphene represents an interesting example where this standard computation scheme had to be redone from scratch \cite{Muller:2008,Muller:2009,Narozhny:2015,Principi:2016,Lucas2016}. An even more dramatic example is given by strongly correlated electron liquids \cite{Spivak:RMP}, where the effects of interactions are nonperturbative, and thus a Boltzmann-like description may not be applicable. Yet the hydrodynamic picture still holds \cite{Andreev:2011} and has to be viewed as a phenomenology that enables one to express various transport observables in terms of pristine kinetic coefficients of the electron liquid and certain thermodynamic quantities. This is our motivation to consider a hydrodynamic description of transport for strongly correlated electron liquids where we do not assume Fermi liquid-like behavior. We also do not assume Galilean invariance to be present. In this study we focus on magnetotransport and frictional drag transresistance in bilayers.  

\textit{Hydrodynamic formalism}. The general linearized set of equations that govern nonrelativistic magnetohydrodynamic transport in two dimensional charged fluids are given by \cite{Hartnoll2007, Foster:2009, Lucas:2015pxa, Lucas2015, Hartnoll2016} (i) the force equations (repeated indices imply summation throughout this work)
\begin{align}
&\partial_t(Mv_i) + \partial_j \mathcal{T}_{ij} = QE_i + S\xi_i + B\epsilon_{ij}J_j, \nonumber \\
&\mathcal{T}_{ij} = P \delta_{ij} - \eta(\partial_i v_j + \partial_j v_i)  - (\zeta - \eta) \partial_k v_k \delta_{ij},
\label{force}
\end{align}
which relate the rate of change of the momentum density to pressure, viscous, thermoelectric and Lorentz forces. $M$ serves as an effective ``mass density" and $Q$ is the effective charge density of the fluid. $\eta$ and $\zeta$ respectively are the shear and bulk viscosities. $E_i$ and $\xi_i$ represent the electric field and thermal gradient. Fluctuations in the fluid pressure $P$ are given by $dP = Qd\mu + SdT$, where $S$ is the entropy density and $\mu$ is the local screened chemical potential per unit charge. (ii) The equations for charge and heat currents read respectively as 
\begin{align}
&J_i = Q v_i - \sigma^Q_{ij}(\partial_j \mu - E_j - B\epsilon_{jk}v_k) - \alpha^Q_{ij} (\partial_j T-\xi_j),  \nonumber \\ 
&J_i^H = TSv_i - T\bar{\alpha}^Q_{ij}(\partial_j \mu - E_j - B\epsilon_{jk}v_k) - \bar{\kappa}^Q_{ij}(\partial_j T-\xi_j), 
\end{align}
where $\sigma^Q$, $\alpha^Q$ and $\bar{\kappa}^Q$ are microscopic ``incoherent" conductivities \cite{Davison:2015taa}, and (iii) the continuity equations are
\begin{equation}
\partial_t Q + \partial_i J_i = \partial_t (TS) + \partial_i J^H_i = 0. 
\end{equation}
Onsager reciprocity requires $\alpha^Q_{ij}(B)=\bar{\alpha}^Q_{ji}(-B)$. The incoherent conductivities, viscosities and thermodynamic properties are derived from correlation functions of the underlying microscopic field theory of the non-Fermi liquid~\cite{Eberlein2016,Eberlein2017,Patel2015}. This is a generalization of the usual theory of hydrodynamics to systems without Galilean invariance.

\textit{Magnetotransport in a single layer}. We consider the steady state solutions of these equations in the presence of a disordered chemical potential $\mu(\mathbf{x})$. In the absence of applied electric fields and temperature gradients, we can apply a background electric field $\bar{E}_i = \partial_i\mu$ to nullify currents and fluid motion, assuming a uniform temperature. We then look for steady state solutions when this background is perturbed by an infinitesimal uniform electric field $\delta E_i$ in linear response~\cite{Lucas2015, Lucas2016}. The difference between the unperturbed and perturbed set of equations gives
\begin{align}
&\partial_i J_i = \partial_i(Q  v_i - \sigma^Q_{ij}(\partial_j \delta \mu - \delta E_j - B\epsilon_{jk} v_k) - \alpha^Q_{ij} \partial_j \delta T) \nonumber \\ 
&=\!\partial_i(TS  v_i - T\bar{\alpha}^Q_{ij}(\partial_j  \delta \mu - \delta E_j - B\epsilon_{jk} v_k)\! - \bar{\kappa}^Q_{ij}\partial_j  \delta T)\! =\! 0, \nonumber \\
&Q(\partial_i\delta\mu -\delta E_i) + S\partial_i\delta T - \partial_j(\eta(\partial_i  v_j + \partial_j  v_i))  \nonumber \\
&-\partial_i((\zeta - \eta) \partial_k  v_k)  = B\epsilon_{ij} J_j,
\label{lheqs}
\end{align}
where the delta-quantities represent deviations from the background values generated by the applied electric field ($v_i\sim O(\delta)$). We read off transport coefficients by looking at the change of the uniform components of their respective currents with respect to the applied electric field. For example, $\sigma_{xx} = \delta J_x(q=0)/\delta E_x$ and $\sigma_{yx} = \delta J_y(q=0)/\delta E_x$. The equations~(\ref{lheqs}) need to be given periodic boundary conditions in order to ensure a unique solution; otherwise one may shift $v$ by a constant and cancel the effects by appropriately shifting $\delta\mu$ by a function that has a constant gradient~\cite{Lucas2015}. We can consider the solution of~(\ref{lheqs}) while treating disorder perturbatively~\cite{Lucas2015, Lucas2016}. Against a uniform background chemical potential, it is easy to see that the only response is a finite uniform velocity field $v_i^{(0)} = \epsilon_{ij} \delta E_j/B$, which implies $\sigma_{ij}^{(0)} = \epsilon_{ij}Q^{(0)}/B$, where $Q^{(0)}$ is the uniform charge density in the absence of any disorder. Introducing a small parameter $\epsilon$ to parameterize the strength of the disorder, we expand $\mu(\mathbf{x}) = \sum_{n=1}^\infty \epsilon^n\mu^{(n)}(\mathbf{x})$. All responses, densities, viscosities and microscopic conductivities may also be expanded in powers of $\epsilon$. For example $v_i(\mathbf{x}) = \sum_{n=0}^\infty \epsilon^n v_i^{(n)}(\mathbf{x})$ and  $Q(\mathbf{x}) = \sum_{n=0}^{\infty}\epsilon^n Q^{(n)}(\mathbf{x})$.

Order by order in $\epsilon$, there are 4 unknowns $\delta\mu^{(n)}$, $\delta T^{(n)}$, $v_i^{(n)}$ and 4 equations in (\ref{lheqs}), so a unique solution is possible. This expansion in disorder strength while keeping $B$ finite implies the assumption that the magnetic field relaxes momentum faster than the disordered potential (see ~\cite{Lucas:2016omy} for when both relaxation rates are comparable). 
The expression for the uniform charge current at $\mathcal{O}(\epsilon^2)$ is (in momentum space)
\begin{align}
&B J_i^{(2)}(k=0) = \epsilon_{ij}Q^{(2)}(k=0)E_j  -i\epsilon_{ij} \nonumber \\
&\times\int_\mathbf{k}~(Q^{(1)}(-\mathbf{k})\delta\mu^{(1)}(\mathbf{k})+S^{(1)}(-\mathbf{k})\delta T^{(1)}(\mathbf{k}))k_j.
\label{j2}
\end{align}
Thus, solving the equations at $\mathcal{O}(\epsilon)$ gives all the information needed to obtain the uniform conductivities up to $\mathcal{O}(\epsilon^2)$.

In general the densities, viscosities and incoherent conductivities depend on $B$, and their functional forms can be deduced from the underlying quantum critical theory, which is beyond the scope of hydrodynamics. However, for small values of $B$ these dependences can be neglected as the dominant effect on magnetoresistance arises from the long-range modulations of the equilibrium density (see Refs. \cite{Grozdanov:2016tdf,Hernandez:2017mch,Baumgartner:2017kme} for other large $B$ effects). This contribution exceeds the one due to the $B$-dependence of the kinetic coefficients of the liquid by a parametrically large factor controlled by the ratio of disorder wavelength to electron equilibration length. We hence set the off-diagonal components of the quantum critical transport to zero. We assume that $\partial Q/\partial\mu\neq 0$ and $\partial S/\partial\mu\neq 0$, so $Q^{(1)}\neq0$ and $S^{(1)}\neq0$. The solutions of~(\ref{lheqs}) are provided in the supplementary material.  

Using~(\ref{j2}) to read off the uniform charge current, we see that $\sigma_{xx,yy}$ are $\mathcal{O}(\epsilon^2)$, whereas $\sigma_{xy}$ is $\mathcal{O}(1)$. Hence the  symmetrized electrical resistance is given by $\mathrm{Tr}~\rho \approx (\sigma_{xx}+\sigma_{yy})/\sigma_{xy}^2$.
This is in general a very complicated function, with a potentially complicated temperature dependence due to the temperature dependences of all the microscopic coefficients. However, if we assume that the disorder is very long wavelength, thus retaining only the leading contribution in the inverse disorder wavelength in the diagonal conductivity, we find a rather simple result
\begin{align}
&\sigma_{ij} = \sigma_{ij}^{(0)} + \epsilon^2\sigma_{ij}^{(2)} =  \epsilon_{ij}\frac{Q^{(0)}+\epsilon^2Q^{(2)}(k = 0)}{B} \nonumber \\
&+ \frac{\epsilon^2}{\eta^{(0)}} \epsilon_{il}\epsilon_{jm}\int_{\mathbf{k}}\frac{k_l k_m|Q^{(1)}(\mathbf{k})|^2}{k^4},
\label{sigmaperturb}
\end{align}
which is consistent with Onsager reciprocity $\sigma_{ij}(B) = \sigma_{ji}(-B)$. All corrections from the microscopic incoherent conductivities appear at higher orders in the inverse disorder wavelength (for details see supplementary information). For the second term of~(\ref{sigmaperturb}) to be smaller than the first, so the perturbative structure is consistent, we must have  
$(\partial Q/\partial\ln\mu) \ll (\eta^{(0)}Q^{(0)})/(\lambda_\mu^4 B))^{1/2}$, where $\lambda_\mu$ is a characteristic wavelength of the disorder. To leading order in $\epsilon$, one gets the symmetrized magnetoresistance at leading order in the inverse disorder wavelength
\begin{align}
&\mathrm{Tr}~\rho(B)-\mathrm{Tr}~\rho(0) =  \frac{\epsilon^2B^2(\partial Q/\partial \mu)_T^2}{Q^{(0)2}\eta^{(0)}} \int_{\mathbf{k}}\frac{|\mu^{(1)}(\mathbf{k})|^2}{k^2} . 
\label{mr}
\end{align}
The temperature dependence of the magnetoresistance is controlled only by the viscosity in this long-wavelength disorder limit, as was the case in ~\cite{Levchenko2017} for the special case of Galilean-invariant fluids ($\sigma^Q=\alpha^Q=0$). However, there the magnetoresistance was controlled only by the viscosity regardless of the spectrum of the disorder. Since we do not expect most non-Fermi liquid metals to be Galilean-invariant, this is an important strengthening of the previous result. It can additionally be shown that the long-wavelength disorder result (\ref{mr}) is also insensitive to the Hall viscosity~\cite{Avron1995} and vorticity susceptibility~\cite{Jensen2012}, which are new parity-odd microscopic transport coefficients that can appear in the presence of a magnetic field.

The above result could enable the extraction of the temperature dependence of the viscosity of the electron liquid as $\delta\rho(B)/\rho(0)\propto 1/\eta^{(0)}$ and thus allow for testing theoretical models of potential non-Fermi liquid states in the hydrodynamic regime. In the supplementary material we also provide results for the magnetothermal resistance. 
 
\textit{Drag transport in bilayers}. For drag type transport \cite{Levchenko:RMP}, we use our hydrodynamic equations for each layer of the bilayer system, with $E=B=\xi=0$. Drag is generated by intrinsic hydrodynamic fluctuations encoded in fluctuating noise terms~\cite{Landau1957,Kovtun2012,Apostolov2014,Chen:2015} added to $T_{ij},J_i,J^H_i$ that are uncorrelated between the layers ($T^{1,2}_{ij} \rightarrow T^{1,2}_{ij} + s^{1,2}_{ij}$,~~$J^{1,2}_i \rightarrow J^{1,2}_i + r^{1,2}_i$,~~$J^{H~1,2}_i \rightarrow J^{H~1,2}_i + g^{1,2}_i$)
\begin{align}
\label{noisecorrel}
&\langle s^{1,2}_{ij}(\mathbf{k},\omega) s^{1,2}_{lm}(\mathbf{k}^\prime,\omega^\prime)\rangle = 2T(\eta^{(0)}(\delta_{il}\delta_{jm}+\delta_{im}\delta_{jl}) \nonumber \\
&+(\zeta^{(0)}-\eta^{(0)})\delta_{ij}\delta_{lm})\delta(\mathbf{k}+\mathbf{k}^\prime)\delta(\omega+\omega^\prime), \\
&\langle r^{1,2}_i(\mathbf{k},\omega) r^{1,2}_j(\mathbf{k}^\prime,\omega^\prime)\rangle = 2T\sigma^{Q(0)}\delta_{ij}\delta(\mathbf{k}+\mathbf{k}^\prime)\delta(\omega+\omega^\prime), \nonumber \\
&\langle g^{1,2}_i(\mathbf{k},\omega) g^{1,2}_j(\mathbf{k}^\prime,\omega^\prime)\rangle = 2T^2\bar{\kappa}^{Q(0)}\delta_{ij}\delta(\mathbf{k}+\mathbf{k}^\prime)\delta(\omega+\omega^\prime), \nonumber \\
&\langle r^{1,2}_i(\mathbf{k},\omega) g^{1,2}_j(\mathbf{k}^\prime,\omega^\prime)\rangle = 2T^2\alpha^{Q(0)}\delta_{ij}\delta(\mathbf{k}+\mathbf{k}^\prime)\delta(\omega+\omega^\prime) \nonumber,
\end{align}
with all other correlators of the sources being zero. These fluctuations induce fluctuations in the charge and entropy densities in the layers ($Q^{(0)}_{1,2}, S^{(0)}_{1,2} \rightarrow Q^{(0)}_{1,2} + \delta Q_{1,2}, S^{(0)}_{1,2}+ \delta S_{1,2}$). The fluctuations of chemical potential and temperature are expressed in terms of the charge and entropy fluctuations 
\begin{equation}
\delta\mu_{1,2} = \left(\frac{\partial\mu}{\partial Q}\right)_S\delta Q_{1,2}+\left(\frac{\partial\mu}{\partial S}\right)_Q\delta S_{1,2},
\end{equation}
and likewise for $\delta T_{1,2}$. We must add to the pressure term in each layer the effects of intra and inter-layer Coulomb forces generated by the fluctuations in the charge densities (the layers are separated by a distance $d$)
\begin{equation}
\delta P_{1,2} \rightarrow \delta P_{1,2} + \frac{2\pi Q^{(0)}}{k}(\delta Q_{1,2} + e^{-k d}\delta Q_{2,1}).
\end{equation}
The drag resistance measures the sensitivity of the electric field induced by the dragging force in the open-circuit passive layer to the current flowing in the driven layer. It is given by
\begin{align}
&\rho_D \equiv \frac{E_{2D}}{J_1} =  \frac{F_{12}(v_x)-F_{12}(0)}{Q^{(0)2}v_x}, \\
&F_{12}(v_x) = -i\int_{\mathbf{k},\omega}\frac{2\pi}{k}k_x e^{-k d} \langle \delta Q_1(\mathbf{k},\omega) \delta Q_2(-\mathbf{k},-\omega) \rangle. \nonumber
\end{align}
The derivation of these force and pressure relations only requires a straightforward application of Coulomb's law. In additional to the noise sources, we also linearize in the velocity $v_x$ (the driven layer is driven by this uniform velocity field, not by an electric field). Note that $J_1 = Q^{(0)} v_x$ is valid even for non-Galilean invariant fluids as the noise terms themselves cannot induce any uniform current flow due to averaged inversion and time-reversal symmetries. Thus $J_1$ must vanish when $v_x=0$, and renormalizations of $Q^{(0)}$ due to the noise terms are subleading.

We neglect the effects of thermal currents: they produce only subleading effects at large spatial separations (see supplementary information for further details). Switching to the basis defined by $\delta Q_\pm = \delta Q_1 \pm \delta Q_2$, $s^\pm_{ij} = s^1_{ij} \pm s^2_{ij}$, $r^\pm_i = r^1_i \pm r^2_i$, the hydrodynamic equations can be reduced to the form 
\begin{align}
&\Pi_\pm \delta Q_\pm + \frac{M}{2}k_x(i(D_\sigma+D_\eta)k^2+\omega) v_x(\delta Q_+ + \delta Q_-) = \nonumber \\
&\frac{M}{2} k_x v_x k_i(r_i^++r_i^-)- i M(D_\eta k^2-i\omega) k_i r^\pm_i - Q^{(0)}k_i s_{ij}^\pm k_j, \nonumber \\
&\Pi_\pm = M(D_\eta k^2 - i\omega)(D_\sigma k^2 - i\omega) \nonumber \\
&+ k Q^{(0)}(2\pi Q^{(0)}(1\pm e^{-kd}) + b_1 k),
\label{pmeqs}
\end{align}
where $D_\sigma=\sigma^{Q(0)}(\partial\mu/\partial Q)_S$, $D_\eta=(\eta^{(0)}+\zeta^{(0)})/M$ and $b_1 = Q^{(0)}(\partial\mu/\partial Q)_S$. The solutions to these equations are linearized in $v_x$: $\delta Q_\pm = \delta Q_\pm^{(0)} + \delta Q_\pm^{(1)} v_x$. Since the $v_x$-less configuration obeys averaged inversion and time-reversal symmetry and $v_x$ always appears as $v_x k_x$ which is odd under inversion, $\delta Q_\pm^{(0)}$ is even under $\mathbf{k} \rightarrow -\mathbf{k}$ whereas $\delta Q_\pm^{(1)}$ is odd. The dragging force may be written as
\begin{align}
&\frac{F_{12}(v_x)\!-\!F_{12}(0)}{i\pi} =\!\! \int_{\mathbf{k},\omega}\frac{k_xv_x}{ke^{kd}}(\langle \delta Q_+^{(0)}(\mathbf{k},\omega) \delta Q_-^{(1)}(-\mathbf{k},-\omega) \rangle \nonumber \\
&-\langle \delta Q_-^{(0)}(\mathbf{k},\omega) \delta Q_+^{(1)}(-\mathbf{k},-\omega) \rangle).
\end{align}
All other terms vanish upon momentum/frequency integration due to even/odd cancellations. Inserting the solutions of (\ref{pmeqs}), we obtain $\rho_D = \rho_D^\sigma + \rho_D^\eta$, where $\rho_D^\sigma$ is generated by the charge fluctuations $r^\pm$ and $\rho_D^\eta$ is generated by the viscous fluctuations $s^\pm$:
\begin{align}
&\rho_D^\sigma = M^3 T\sigma^{Q(0)} \int_0^\infty dk \int_\omega \frac{(D_\sigma+D_\eta)k^7(\omega^2 -D_\eta^2 k^4)}{e^{2kd}|\Pi_+|^2|\Pi_-|^2}, \nonumber \\
&\rho_D^\eta = M Q^{(0)2}T(\eta^{(0)}+\zeta^{(0)}) \int_0^\infty dk \int_\omega \frac{(D_\sigma+D_\eta)k^9}{e^{2kd}|\Pi_+|^2|\Pi_-|^2}. 
\label{rhoDpm} 
\end{align}

This yields a complicated integral expression for $\rho_D$. We can however make simplifications in the regimes of ``large" and ``small" $d$. The model of Fermi surface coupled to $U(1)$ gauge field has roughly the following properties~\cite{Halperin1993,Eberlein2016,Eberlein2017} for dynamical critical exponent $z=3$ ($m$ is the effective fermion mass), corresponding to the case of short-ranged interactions of composite fermions~\cite{Halperin1993}:
\begin{align}
&Q^{(0)} \propto e k_F^2,~~M \propto m k_F^2,~~\left(\frac{\partial\mu}{\partial Q}\right)_S \propto \frac{\hbar^2}{e^2m},~~b_1 \propto \frac{E_F}{e}, \nonumber \\
&\sigma^{Q(0)} \propto \left(\frac{E_F}{k_BT}\right)^{2/3}\frac{e^2}{\hbar},~~~\eta^{(0)} \sim \zeta^{(0)} \propto \left(\frac{E_F}{k_BT}\right)^{2/3}\hbar k_F^2.
\label{fscr}
\end{align}
$d$ is said to be ``large" when $d^3 \gg d_c^3 \equiv M(D_\eta+D_\sigma)^2/Q^{(0)2}$. This gives
\begin{equation}
(k_F d)^3 \gg \left(\frac{E_F}{k_BT}\right)^{4/3} \frac{\varepsilon E_F}{e^2k_F}.
\end{equation}
We have set the electrostatic permittivity $\varepsilon = 1$ so far in the paper but restored it in the last equation. We also demand $d \gg d_e \equiv \hbar^2\varepsilon/(e^2m)$,
which is trivially achieved as $d_e$ is typically a very small distance scale ($\mathcal{O}(10^{-10})$~m for $m\sim m_e/4$).  

For $d\gg d_c$ we obtain the leading contributions 
\begin{align}
&\rho_D^\sigma  \sim \frac{\hbar}{e^2} \left(\frac{k_B T}{E_F}\right)^{1/3}\frac{\ln^4\left(\frac{dk_F^{2/3}}{d_e^{1/3}}\left(\frac{k_BT}{E_F}\right)^{4/9}\right)}{(k_Fd)^4}, \nonumber \\
&\rho_D^\eta \sim \frac{\hbar}{e^2} \left(\frac{k_BT}{E_F}\right)^{1/3} \frac{\ln^5\left(\frac{dk_F^{2/3}}{d_e^{1/3}}\left(\frac{k_BT}{E_F}\right)^{4/9}\right)}{(k_F d)^5/(k_F d_e)}.
\label{rhoDfinal}
\end{align}
$\rho_D^\sigma$ and $\rho_D^\eta$ have the same temperature scaling up to logarithms. However, $\rho_D^\eta$ falls off faster with $d$ than $\rho_D^\sigma$. This results should be contrasted to that obtained earlier for Fermi liquids \cite{Apostolov2014}. Note that even though the power dependence on temperature is $T^{1/3}$, there is a $\ln^4(T)$ correction, which will make the temperature dependence appear faster than $T^{1/3}$ but slower than $T$, which is consistent with the data of Refs.~\cite{Lilly1998,Jorger2000} at large separations.

\begin{figure}
\begin{center}
\includegraphics[width=\columnwidth]{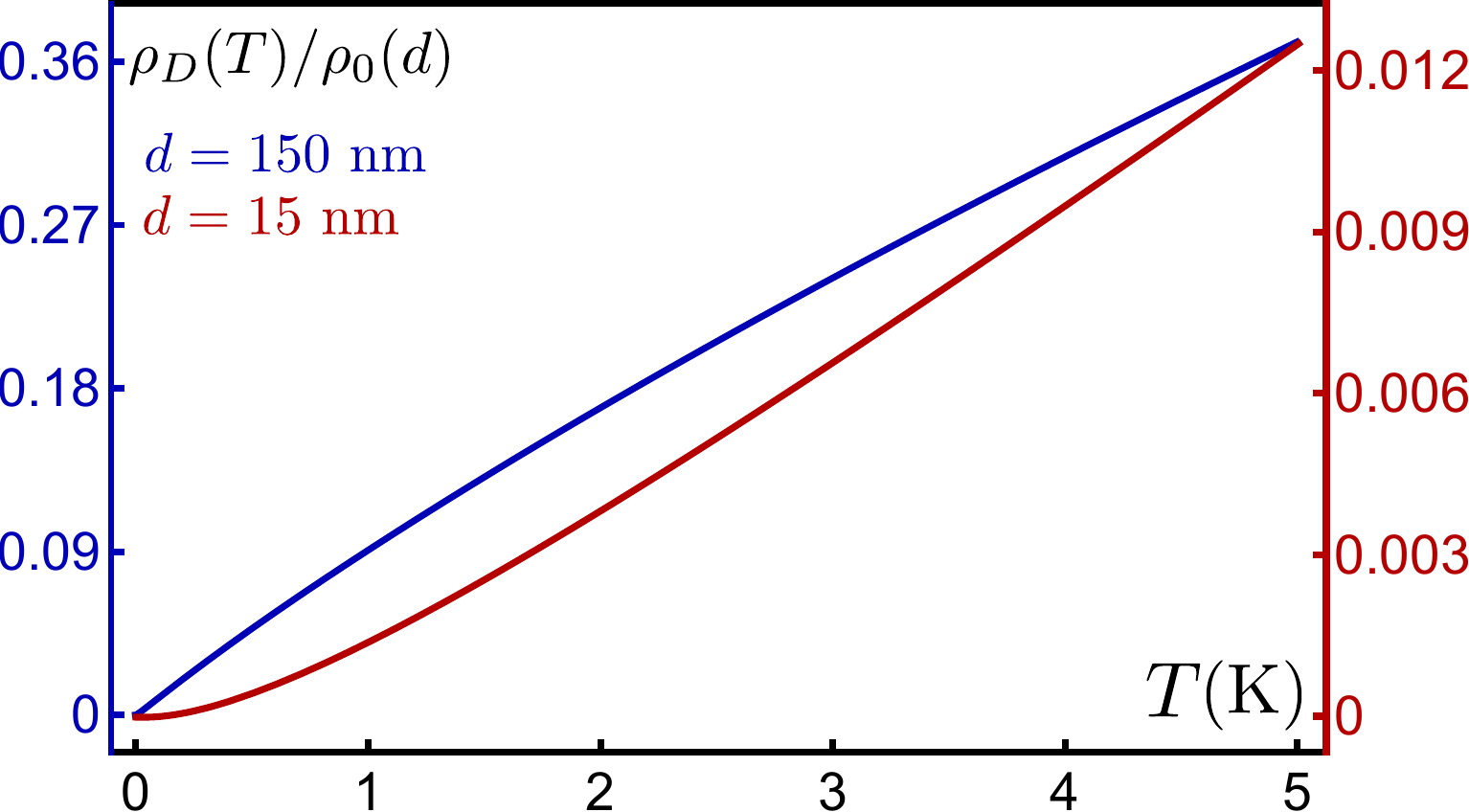}
\end{center}
\caption{Normalized drag resistance $\rho_D(T)/\rho_0(d)$. $\rho_D(T)=\rho_D^\sigma(T)+\rho_D^\eta(T)$ is obtained by numerically evaluating~(\ref{rhoDpm}) for two different spatial separations. $\rho_0(d)=(\hbar/e^2)/(k_Fd)^4$.  Note the crossover from positive to negative curvature as $d$ is increased. This feature holds for other values of the dynamical critical exponent $2<z<3$ as well that can appear in the theory of~\cite{Halperin1993}. We use $T_F\sim 50$~K and $m\sim m_e/4$ ($d_c \sim 10$~nm at $T \sim 5$~K). We set all constants of proportionality in (\ref{fscr}) to $1$. Numerical values should be treated as order-of-magnitude estimates only.}
\label{rhoDscaling}
\end{figure}

At small separations $d\ll d_c$, all contributions to $\rho_D$ scale as $T^{n>1}$ (see further details in supplementary materials). This is again consistent with ~\cite{Jorger2000}, which shows an apparent crossover from positive to negative curvature in $\rho_D(T)$ as a function of $T$ as $d$ is increased. In Fig. \ref{rhoDscaling} we show $\rho_D(T)$ obtained by numerically evaluating the integrals without the above approximations that confirm the qualitative behaviors we discussed. It should be carefully noted that in Fig. \ref{rhoDscaling} the line corresponding to $d=150$ nm appears superficially above the line of $d=15$ nm plot which is due to the choice of the normalization factor $\rho_0(d)$. Drag is obviously a decaying function of inter-layer separation $d$ as is clear from \eqref{rhoDfinal}. 

\textit{Discussion}. The most extensively studied example of transresistance in the case of non-Fermi liquids corresponds to inter-layer frictional Coulomb drag between bilayers of half-filled Landau levels \cite{Jorger2000,Lilly1998,Zelakiewicz:2000,Price:2010}. The theoretical approach that has proved most useful for understanding the filling fraction $\nu=1/2$ state is the fermion Chern-Simons field theory, which is based in turn on the composite-fermion picture \cite{Halperin1993}. Previous calculations \cite{Ussishkin:1997,Sakhi:1997,Kim:1999} showed that the dominant low-temperature behavior for $\rho_D$ scales with temperature as $T^{4/3}$ (see supplementary material for a brief summary of this result). This unique power exponent can be traced back to a special momentum dependence of the electronic longitudinal conductivity, as can be deduced from surface acoustic wave measurements. Indeed, in the composite-fermion picture, at $\nu=1/2$, the density response at small frequencies and small wave-vectors is of the form $\propto (k^3-8\pi i\chi\omega k_F)^{-1}$, which can be viewed as slow diffusion with an effective diffusion constant that vanishes linearly with $k$ (where $\chi$ is the thermodynamic compressibility of the $\nu=1/2$ state). Since the typical frequency is set by temperature $\omega\sim T$, the pole structure of long-wavelength density fluctuations sets a characteristic scale for momentum transfer between the layers $k\propto T^{1/3}$ that then carriers over to drag resistance $\rho_D\propto T^{4/3}$. This should be contrasted the Fermi liquid prediction $\rho_D\propto T^2$ at lowest temperatures, and our prediction $\rho_D\propto T^{1/3}$. In our current understanding, the results of  \cite{Ussishkin:1997} correspond to the ``collisionless" regime of transport with respect to intra-layer collisions, namely a long equilibration length as mediated by interactions with the gauge field. We considered the opposite collision-dominated regime where this length scale is assumed to be short. This should explain the difference between the power exponents $4/3$ and $1/3$ between two limiting cases. We hope that understanding different transport regimes and corresponding temperature dependencies will be of help for the interpretation of future experiments, as it also deepens our current understanding of the existing transport data and corresponding theories.     

\textit{Acknowledgements}. We thank A. Andreev, L. V. Delacretaz, B. Halperin, P. Kim, A. Lucas, and S. Sachdev for helpful discussions. A.A.P. acknowledges support by NSF Grant DMR-1360789. R.A.D. is supported by the Gordon and Betty Moore Foundation Grant GBMF-4306. The work of A.L. was financially supported in part by NSF Grant DMR-1653661, and by the Office of the Vice Chancellor for Research and Graduate Education with funding from the Wisconsin Alumni Research Foundation.

\bibliography{hydro}

\appendix
\onecolumngrid
\section{\large Supplementary information}

\section{Solution to linearized disordered magneto-hydrodynamic equations}

At $\mathcal{O}(\epsilon)$ we get the following equations in momentum space from (\ref{lheqs}) of the main text
\begin{align}
&-Q^{(1)}\delta E_i + ik_i\left(Q^{(0)}\delta\mu^{(1)}+S^{(0)}\delta T^{(1)} - i\zeta^{(0)}k_j v_j^{(1)}\right)  +\eta^{(0)}k^2 v_i^{(1)} = B \epsilon_{ij} J_j^{(1)},  \\
&J_i^{(1)} = Q^{(0)} v_i^{(1)} + Q^{(1)}\epsilon_{ij}\frac{\delta E_j}{B}  -\sigma^{Q(0)}_{ij}(ik_j\delta\mu^{(1)}-B\epsilon_{jk} v_k^{(1)}) -i\alpha^{Q(0)}_{ij}k_j\delta T^{(1)},  \\
&J_i^{H(1)} = TS^{(0)} v_i^{(1)} + TS^{(1)}\epsilon_{ij}\frac{\delta E_j}{B} -T\bar{\alpha}^{Q(0)}_{ij}(ik_j\delta\mu^{(1)}-B\epsilon_{jk} v_k^{(1)})-i\bar{\kappa}^{Q(0)}_{ij}k_j\delta T^{(1)},  \\
&k_i J_i^{(1)} = k_i J_i^{H(1)} = 0.
\label{lheqse1}
\end{align}
The zero-momentum limit of the equations~(\ref{lheqse1}) implies that $v_i^{(1)}$ and $J_i^{(1)}$ have only finite momentum components, since the periodic boundary conditions ensure that $\lim_{k\rightarrow0}k_i\delta\mu = \lim_{k\rightarrow0}k_i\delta T = 0$. $Q^{(1)}$ also has only finite momentum components and a solution to $Q^{(0)}v_i^{(1)}(k=0) = -B\sigma_{ij}^{Q(0)}\epsilon_{jk}v_{k}^{(1)}(k=0)$ doesn't exist. Hence we need to look at $\mathcal{O}(\epsilon^2)$ to obtain nontrivial uniform conductivities.
 
Using the first equation of~(\ref{lheqse1}) to read off $J_i^{(1)}$, we obtain
\begin{equation}
k_i J_i^{(1)} = -\frac{\epsilon_{ij}}{B}k_i(\eta^{(0)}k^2v_j^{(1)}-Q^{(1)}\delta E_j).
\end{equation}
Note that it is impossible to set the right hand side of this equation to zero in the limit of vanishing shear viscosity $\eta^{(0)}$. Thus, this perturbative expansion is valid only for finite background shear viscosities.

The solution to (\ref{lheqse1}) is given by
\begin{align}
&\delta\mu^{(1)} = i \epsilon_{ij}E_i k_j  \Bigg(B^4 Q^{(1)}\sigma^{Q(0)} \left(\bar{\kappa}^{Q(0)}\sigma^{Q(0)}-\alpha^{Q(0)2} T\right)+ B^2 \Big(\alpha^{Q(0)} Q^{(1)} T \left(-\alpha^{Q(0)} k^2 (\eta^{(0)}+\zeta^{(0)})-2 Q^{(0)} S^{(0)}\right) \nonumber \\
&+\sigma^{Q(0)} \left((\zeta^{(0)}+2\eta^{(0)})\bar{\kappa}^{Q(0)} k^2 Q^{(1)}+\alpha^{Q(0)}\eta^{(0)} k^2 (-S^{(1)}) T+Q^{(1)} S^{(0)2} T\right)+\bar{\kappa}^{Q(0)} Q^{(0)2} Q^{(1)}\Big) \nonumber \\
&+\eta^{(0)} k^2 \left(S^{(1)} T \left(-\alpha^{Q(0)} k^2 (\eta^{(0)}+\zeta^{(0)})-Q^{(0)} S^{(0)}\right)+\bar{\kappa}^{Q(0)} k^2 Q^{(1)} (\zeta^{(0)}+\eta^{(0)})+Q^{(1)} S^{(0)2} T\right)\Bigg) \nonumber \\
&\times\Bigg(B\eta^{(0)} k^4 \Big(B^2\bar{\kappa}^{Q(0)}\sigma^{Q(0)2}+\sigma^{Q(0)} \left(T (\alpha^{Q(0)} B+S^{(0)}) (S^{(0)}-\alpha^{Q(0)} B)+\bar{\kappa}^{Q(0)} k^2 (\zeta^{(0)}+\eta^{(0)})\right) \nonumber \\
&-\alpha^{Q(0)2} k^2 T (\zeta^{(0)}+\eta^{(0)})+\bar{\kappa}^{Q(0)} Q^{(0)2}-2\alpha^{Q(0)} Q^{(0)} S^{(0)} T\Big)\Bigg)^{-1},  \\
&\delta T^{(1)} = -i \epsilon_{ij}E_i k_j T \Bigg((\alpha^{Q(0)} Q^{(1)}-S^{(1)} \sigma^{Q(0)}) \left(B^2 \sigma^{Q(0)}+k^2 (\zeta^{(0)}+\eta^{(0)})\right)-Q^{(0)2} S^{(1)} +Q^{(0)} Q^{(1)} S^{(0)}\Bigg) \nonumber \\
&\times\Bigg(B k^2 \Big(B^2 \bar{\kappa}^{Q(0)} \sigma^{Q(0)2}+\sigma^{Q(0)} \left(T (\alpha^{Q(0)} B+S^{(0)}) (S^{(0)}-\alpha^{Q(0)} B)+\bar{\kappa}^{Q(0)} k^2 (\zeta^{(0)}+\eta^{(0)})\right) \nonumber \\
&-\alpha^{Q(0)2} k^2 T (\zeta^{(0)}+\eta^{(0)})+\bar{\kappa}^{Q(0)} Q^{(0)2}-2 \alpha^{Q(0)} Q^{(0)} S^{(0)} T\Big)\Bigg)^{-1}, 
\end{align}
\begin{align}
&v_i^{(1)} = \frac{\epsilon_{lm}E_l k_m}{k^4} \Bigg(\frac{\epsilon_{ij}k_j Q^{(1)}}{\eta^{(0)}} -\Big(k^2 k_i(-\bar{\kappa}^{Q(0)} Q^{(0)} Q^{(1)}+\alpha^{Q(0)} T (Q^{(0)} S^{(1)}+Q^{(1)} S^{(0)})-S^{(0)} S^{(1)} \sigma^{Q(0)} T)\Big) \nonumber \\
&\times\Big(B \Big(B^2 \bar{\kappa}^{Q(0)} \sigma^{Q(0)2}+\sigma^{Q(0)} \left(T (\alpha^{Q(0)} B+S^{(0)}) (S^{(0)}-\alpha^{Q(0)} B)+\bar{\kappa}^{Q(0)} k^2 (\zeta^{(0)}+\eta^{(0)})\right) \nonumber \\
&-\alpha^{Q(0)2} k^2 T (\zeta^{(0)}+\eta^{(0)})+\bar{\kappa}^{Q(0)} Q^{(0)2}-2 \alpha^{Q(0)} Q^{(0)} S^{(0)} T\Big)\Big)^{-1}\Bigg).
\label{oeps}
\end{align}

Using (\ref{j2}) of the main text, the expression for the uniform electrical conductivity $\sigma_{ij}$ retaining only the leading and next-to-leading contributions in the inverse disorder wavelength in the diagonal conductivity is given by
\begin{align}
&\sigma_{ij} = \sigma_{ij}^{(0)} + \epsilon^2\sigma_{ij}^{(2)} =  \epsilon_{ij}\frac{Q^{(0)}+\epsilon^2Q^{(2)}(k = 0)}{B} + \frac{\epsilon^2}{\eta^{(0)}} \epsilon_{il}\epsilon_{jm}\int_{\mathbf{k}}\frac{k_l k_m|Q^{(1)}(\mathbf{k})|^2}{k^4} \nonumber \\
&+ \frac{\epsilon^2}{\eta^{(0)}B^2\left(\sigma^{Q(0)} \left(B^2 \left(\bar{\kappa}^{Q(0)} \sigma^{Q(0)}-\alpha^{Q(0)2} T\right)+S^{(0)2} T\right)+\bar{\kappa}^{Q(0)} Q^{(0)2}-2 \alpha^{Q(0)} Q^{(0)} S^{(0)} T\right)} \epsilon_{il}\epsilon_{jm}\int_{\mathbf{k}}\frac{k_lk_m}{k^2} \nonumber \\
&\times \Big[\eta^{(0)}T \left|Q^{(1)}(-\mathbf{k}) S^{(0)}-S^{(1)}(\mathbf{k}) Q^{(0)}\right| ^2 \nonumber \\
&-B^2 Q^{(1)}(\mathbf{k}) \left(-\bar{\kappa}^{Q(0)} \sigma^{Q(0)} (\zeta^{(0)}+2 \eta^{(0)}) Q^{(1)}(-\mathbf{k})+\alpha^{Q(0)2}T (\zeta^{(0)}+\eta^{(0)}) Q^{(1)}(-\mathbf{k})+\alpha^{Q(0)} \eta^{(0)}\sigma^{Q(0)} T S^{(1)}(-\mathbf{k})\right) \nonumber \\
&+B^2 \eta^{(0)}(-T) S^{(1)}(\mathbf{k}) (\sigma^{Q(0)} (\alpha^{Q(0)} Q^{(1)}(-\mathbf{k})-\sigma^{Q(0)} S^{(1)}(-\mathbf{k})))\Big] \nonumber \\
&+\frac{\epsilon^2}{\eta^{(0)}} \frac{(\zeta^{(0)}+\eta^{(0)}) \left(\alpha^{Q(0)2}T-\bar{\kappa}^{Q(0)} \sigma^{Q(0)}\right)}{\sigma^{Q(0)} \left(B^2 \left(\bar{\kappa}^{Q(0)} \sigma^{Q(0)}-\alpha^{Q(0)2} T\right)+S^{(0)2} T\right)+\bar{\kappa}^{Q(0)} Q^{(0)2}-2 \alpha^{Q(0)} Q^{(0)} S^{(0)} T}\epsilon_{il}\epsilon_{jm}\int_{\mathbf{k}}\frac{k_l k_m|Q^{(1)}(\mathbf{k})|^2}{k^2}.
\label{eq:sigmaperturb}
\end{align} 
Note that the leading disorder-induced contribution depends only upon the shear viscosity $\eta^{(0)}$, and that all corrections coming from the microscopic incoherent conductivities occur at higher orders in the inverse disorder wavelength.

In the presence of a magnetic field, the stress tensor $\mathcal{T}_{ij}$ can contain the effects of new parity-odd microscopic transport coefficients. These are the Hall viscosity $\eta_H$~\cite{Avron1995} and the vorticity susceptibility $\chi_\Omega$~\cite{Jensen2012}, which are both proportional to $B$. The stress tensor is modified to
\begin{equation}
\mathcal{T}_{ij} \rightarrow \mathcal{T}_{ij} + \frac{\eta_H}{2}(\epsilon_{ik}(\partial_k v_j + \partial_j v_k) + \epsilon_{jk}(\partial_k v_i + \partial_i v_k)) + \chi_\Omega \delta_{ij} \epsilon_{kl}\partial_k v_l.
\end{equation} 
At $\mathcal{O}(\epsilon)$, this becomes
\begin{equation}
\mathcal{T}_{ij}^{(1)} \rightarrow \mathcal{T}_{ij}^{(1)} + i\frac{\eta_H^{(0)}}{2}(\epsilon_{il}(k_l v_j^{(1)} + k_j v_l^{(1)}) + \epsilon_{jl}(k_l v_i^{(1)} + k_i v_l^{(1)})) + i\chi_\Omega^{(0)} \delta_{ij} \epsilon_{lm}k_l v_m^{(1)}.
\end{equation} 
Then, repeating our solution, we find that the long-wavelength disorder result for the magnetoresistance given by (\ref{mr}) of the main text is unaffected by these terms. 

\section{Magneto-thermal transport in the clean system}

To obtain the thermal resistance, 
\begin{equation}
\mathrm{Tr}~\rho_{\mathrm{th}}= \frac{\kappa_{xx}+\kappa_{yy}}{\kappa_{xx}\kappa_{yy}-\kappa_{xy}\kappa_{yx}},
\end{equation}
we apply a temperature gradient $\partial_i T^{(0)} = -\xi_x\hat{x}$, an electric field $E_x=-S^{(0)}\xi_x/Q^{(0)}$ to block electric currents, and solve the hydrodynamic equations in linear response. The clean solution is 
\begin{align}
&v^{(0)}_x= \xi_x\frac{S^{(0)} \sigma^{Q(0)}-\alpha^{Q(0)} Q^{(0)}}{B^2 \sigma^{Q(0)2}+Q^{(0)2}}, \\
&v^{(0)}_y= \frac{B\sigma^{Q(0)}}{Q^{(0)}}v^{(0)}_x.
\end{align}
This choice gives $J=0$ but $J^H\neq0$. Therefore, there is a finite magneto-thermal resistance in the clean system itself, given by
\begin{align}
&\mathrm{Tr}~\rho_{\mathrm{th}}(B)-\mathrm{Tr}~\rho_{\mathrm{th}}(0) = \frac{2 B^2 \left(Q^{(0)2} T (\alpha^{Q(0)} Q^{(0)}-S^{(0)} \sigma^{Q(0)})^2 \left(\alpha^{Q(0)2} T-\bar{\kappa}^{Q(0)} \sigma^{Q(0)}\right)\right)}{\left(2 \alpha^{Q(0)} Q^{(0)} S^{(0)} T-S^{(0)2} \sigma^{Q(0)} T-\bar{\kappa}^{Q(0)} Q^{(0)2}\right)^3} + \mathcal{O}(B^4).
\label{thmrdiag}
\end{align}

Unlike the charge magneto-transport in the clean system, the magneto-thermal resistance is actually sensitive at the same order in $B$ to the off-diagonal microscopic conductivities $\sigma^{Q(0)}_{xy}=-\sigma^{Q(0)}_{yx}= a_1 B$, $\bar{\kappa}^{Q(0)}_{xy}=-\bar{\kappa}^{Q(0)}_{yx}=a_3 B$ and $\alpha^{Q(0)}_{xy}=\bar{\alpha}^{Q(0)}_{xy}=-\alpha^{Q(0)}_{yx}=-\bar{\alpha}^{Q(0)}_{yx}=a_2B$ at small $B$. Thus (\ref{thmrdiag}) should be modified to
\begin{align}
&\mathrm{Tr}~\rho_{\mathrm{th}}(B)-\mathrm{Tr}~\rho_{\mathrm{th}}(0) = \frac{2 B^2 Q^{(0)2}}{\left(\bar{\kappa}^{Q(0)}  Q^{(0)2}+2 \alpha^{Q(0)}Q^{(0)} S^{(0)} T+\sigma^{Q(0)}  S^{(0)2} T\right)^3} \nonumber \\
&\times \Bigg(a_3^2 Q^{(0)4}-2 a_3 Q^{(0)} T \left(2 a_2 Q^{(0)2} S^{(0)}-a_1 Q^{(0)} S^{(0)2}+(\alpha^{Q(0)}  Q^{(0)}-\sigma^{Q(0)}  S^{(0)})^2\right) \nonumber \\
&+T \Big(2 a_2 (\alpha^{Q(0)}  Q^{(0)}-\sigma^{Q(0)}  S^{(0)}) \left(\bar{\kappa}^{Q(0)} Q^{(0)2}-\sigma^{Q(0)}  S^{(0)2} T\right)+4 a_2^2 Q^{(0)2} S^{(0)2} T-4 a_1 a_2 Q^{(0)} S^{(0)3} T \nonumber \\
&-2 a_1 S^{(0)} (\sigma^{Q(0)}  S^{Q(0)}-\alpha^{Q(0)}  Q^{(0)}) (\alpha^{Q(0)}  S^{(0)} T-\bar{\kappa}^{Q(0)}  Q^{(0)})+a_1^2 S^{(0)4} T \nonumber \\
&+(\alpha^{Q(0)}  Q^{(0)}-\sigma^{Q(0)}  S^{(0)})^2 \left(\alpha^{Q(0)2} T-\bar{\kappa}^{Q(0)}  \sigma^{Q(0)} \right)\Big)\Bigg) +\mathcal{O}(B^4).
\label{thrmfull}
\end{align}

\section{Solution to hydrodynamic equations of the bilayer system}

Since we have a uniform velocity field $v_x$ in layer 1, we modify its force equation from (\ref{force}) of the main text to 
\begin{equation}
\frac{d}{dt}(M_1v_i^1) + \partial_j \mathcal{T}_{ij}^1 = 0,
\label{f1}
\end{equation}
where $d/dt = \partial_t + v^1_j \partial_j$ includes the contribution of converting from a co-moving frame. This is a non-linear effect which is present in both Galilean and relativistic hydrodynamics, for which the full non-linear theories are known.

After linearization in the noise sources and $v_x$, the force equations for the two layers in momentum/frequency space become 
\begin{align}
&iM(v_x k_x -\omega)\delta v_i^{1} - i \omega v_x \delta M_1+ ik_i\left(b_1\delta Q_1 + b_2\delta S_1 + \frac{2\pi}{k}Q^{(0)}(\delta Q_1 + e^{-kd}\delta Q_2)\right) + (\zeta^{(0)} k_i k_j + \eta^{(0)} k^2 \delta_{ij})\delta v_j^{1}  +  i k_j s_{ij}^1 = 0, \nonumber \\
&-iM\omega \delta v_i^{2} + ik_i\left(b_1\delta Q_2 + b_2\delta S_2 + \frac{2\pi}{k}Q^{(0)}(\delta Q_2 + e^{-kd}\delta Q_1)\right) + (\zeta^{(0)} k_i k_j + \eta^{(0)} k^2 \delta_{ij})\delta v_j^{2} +  i k_j s_{ij}^2 = 0,
\end{align}
where
\begin{align}
&b_1 = Q^{(0)} \left(\frac{\partial\mu}{\partial Q}\right)_S + S^{(0)} \left(\frac{\partial T}{\partial Q}\right)_S, \\
&b_2 = Q^{(0)} \left(\frac{\partial\mu}{\partial S}\right)_Q + S^{(0)} \frac{T}{C_V}.
\end{align}
At low temperatures we also expect $\delta M_1 \approx (M/Q^{(0)})\delta Q_1$ in non-Fermi liquids with a Fermi surface (See (\ref{fscr}) of the main text). The above equations then provide
\begin{align}
&k_i \delta v_i^1 = \frac{2\pi k Q^{(0)} (\delta Q_1 + e^{-k d}\delta Q_2) +k^2(b_1 \delta Q_1 + b_2 \delta S_1) - k_x v_x \omega (M/Q^{(0)}) \delta Q_1 + k_i s_{ij}^1 k_j  }{M (iD_\eta k^2 + \omega -k_x v_x)}, \\
&k_i \delta v_i^2 = \frac{2\pi k Q^{(0)} (\delta Q_2 + e^{-k d}\delta Q_1) +k^2(b_1 \delta Q_2 + b_2 \delta S_2) + k_i s_{ij}^2 k_j  }{M (iD_\eta k^2 + \omega)},
\label{kv}
\end{align}
with $D_\eta = (\eta^{(0)}+\zeta^{(0)})/M$. To disable thermal currents, we set $\bar{\kappa}^{Q(0)}=\alpha^{Q(0)}=S^{(0)}=0$. This makes $\delta S_{1,2} = 0$. 
Inserting (\ref{kv}) into the continuity equations for charge, we obtain
\begin{align}
&M(D_\sigma k^2-i\omega)(D_\eta k^2 -i \omega)\delta Q_1 + Q^{(0)} (2\pi k Q^{(0)} (\delta Q_1 + e^{-kd} \delta Q_2) + b_1 k^2 \delta Q_1 + k_i s_{ij}^1 k_j) \nonumber \\
&+ iM((D_\sigma+D_\eta)k^2-i\omega)k_x v_x \delta Q_1 = - i M(D_\eta k^2-i\omega+ik_x v_x) k_i r_i^1, \\
&M(D_\sigma k^2-i\omega)(D_\eta k^2 -i \omega)\delta Q_2 + Q^{(0)} (2\pi k Q^{(0)} (\delta Q_2 + e^{-kd} \delta Q_1) + b_1 k^2 \delta Q_2 + k_i s_{ij}^1 k_j)  = - i M(D_\eta k^2-i\omega) k_i r_i^2, 
\label{nothermal}
\end{align}
with $D_\sigma = \sigma^{Q(0)} (\partial\mu/\partial Q)_S$. In the $\pm$ basis,~(\ref{nothermal}) turns into (\ref{pmeqs}) of the main text.

In the Halperin-Lee-Read (HLR) composite fermion theory of the $\nu=1/2$ quantum Hall state~\cite{Halperin1993}, the composite fermions are also subjected to a transverse magnetic field corresponding to the deviation in their local density from half-filling. Thus, to linear order in the noise terms, we should also add a term $\propto \delta_{iy} \delta Q_1 Q^{(0)} v_x$ to the right hand side of~(\ref{f1}) and shift $J_y^1$ by a term $\propto \sigma^{Q(0)}\delta Q_1 v_x$. However, these terms end up producing parity-odd contributions to $\rho_D$ that vanish upon integration over $k_y$, and can thus be ignored.     

The frequency integrations in (\ref{rhoDpm}) of the main text may be performed first
\begin{align}
&\int_{-\infty}^{\infty} d\omega \frac{4\pi Q^{(0)2}k^3 e^{-kd}M(D_\sigma+D_\eta)}{|\Pi_+|^2|\Pi_-|^2} = \int_{-\infty}^{\infty} d\omega \frac{\omega_0(\omega_+^2-\omega_-^2)}{((\omega^2-\omega_+^2)^2+\omega^2\omega_0^2)((\omega^2-\omega_-^2)^2+\omega^2\omega_0^2)} \nonumber \\
&=\frac{\pi(\omega_+^2-\omega_-^2)(2\omega_0^2+\omega_+^2+\omega_-^2)}{\omega_+^2\omega_-^2((\omega_+^2-\omega_-^2)^2+2\omega_0^2(\omega_+^2+\omega_-^2))}. \\
&\int_{-\infty}^{\infty} d\omega \frac{k^2 \omega^2M(D_\sigma+D_\eta)}{|\Pi_+|^2|\Pi_-|^2} = \frac{2\pi/M}{(\omega_+^2-\omega_-^2)^2+2\omega_0^2(\omega_+^2+\omega_-^2)}, \\
&\omega_\pm^2 = k(D_\sigma D_\eta k^3 M + Q^{(0)}(kb_1+2\pi Q^{(0)}\chi_\pm)), \\
&\omega_0^2 = Mk^4(D_\sigma+D_\eta)^2. 
\end{align}
We then get
\begin{align}
&\rho_D^\sigma =  \int_0^\infty dk \frac{2\pi T \sigma^{Q(0)} M k^3 e^{-2kd}}{16\pi^2Q^{(0)4}e^{-2kd}+4k^3M(D_\eta+D_\sigma)^2(D_\eta D_\sigma k^3 M + Q^{(0)}(k b_1 + 2\pi Q^{(0)}))} \\
&- \int_0^\infty dk \frac{\pi T \sigma^{Q(0)} D_\eta^2M^2k^6 e^{-2kd}(k^3M(D_\eta+D_\sigma)^2+D_\eta D_\sigma k^3 M+Q^{(0)}(k b_1 +2\pi Q^{(0)}))}{(D_\eta D_\sigma k^3 M+Q^{(0)}(k b_1 +2\pi Q^{(0)}))^2-4\pi^2Q^{(0)4}e^{-2kd}} \nonumber \\
&\times\frac{1}{8\pi^2Q^{(0)4}e^{-2kd}+2k^3M(D_\eta+D_\sigma)^2(D_\eta D_\sigma k^3 M+Q^{(0)}(k b_1 +2\pi Q^{(0)}))}, \nonumber \\
&\rho_D^\eta = \int_0^\infty dk \frac{\pi T (\eta^{(0)}+\zeta^{(0)}) Q^{(0)2}k^4 e^{-2kd}(k^3M(D_\eta+D_\sigma)^2+D_\eta D_\sigma k^3 M+Q^{(0)}(k b_1 +2\pi Q^{(0)}))}{(D_\eta D_\sigma k^3 M+Q^{(0)}(k b_1 +2\pi Q^{(0)}))^2-4\pi^2Q^{(0)4}e^{-2kd}} \nonumber \\
&\times\frac{1}{8\pi^2Q^{(0)4}e^{-2kd}+2k^3M(D_\eta+D_\sigma)^2(D_\eta D_\sigma k^3 M+Q^{(0)}(k b_1 +2\pi Q^{(0)}))}.
\label{rhoDintsU}
\end{align}
These expressions can be written in the following scaling form, restoring factors of $k_B$ and $\varepsilon$, ($d_c = (\varepsilon M(D_\eta+D_\sigma)^2/Q^{(0)2})^{1/3}$, $d_e=\hbar^2\varepsilon/(e^2m)$)
\begin{align}
&\rho_D^\sigma =  \frac{k_BT\sigma^{Q(0)} M}{4d^4Q^{(0)4}}\int_0^\infty dp \frac{ p^3 e^{-2p}}{2\pi e^{-2p}+p^3 (d_c/d)^3(1+p d_e/(2\pi d)) + \Delta (p^6/(2\pi)) (d_c/d)^6} \\
&- \frac{k_BT \sigma^{Q(0)} D_\eta^2M^2\varepsilon}{8\pi d^7Q^{(0)6}}\int_0^\infty dp \frac{p^6 e^{-2p}((p^3/(2\pi))(d_c/d)^3(1+\Delta)+1+p d_e/(2\pi d))}{(\Delta (p^3/(2\pi))(d_c/d)^3+1 + pd_e/(2\pi d))^2-e^{-2p}} \nonumber \\
&\times\frac{1}{2\pi e^{-2p}+p^3 (d_c/d)^3(1+p d_e/(2\pi d)) + \Delta (p^6/(2\pi)) (d_c/d)^6}, \nonumber \\
&\rho_D^\eta = \frac{k_BT (\eta^{(0)}+\zeta^{(0)})\varepsilon}{8\pi d^5Q^{(0)4}}\int_0^\infty dp \frac{p^4 e^{-2p}((p^3/(2\pi))(d_c/d)^3(1+\Delta)+1+p d_e/(2\pi d))}{(\Delta (p^3/(2\pi))(d_c/d)^3+1 + pd_e/(2\pi d))^2-e^{-2p}} \nonumber \\
&\times\frac{1}{2\pi e^{-2p}+p^3 (d_c/d)^3(1+p d_e/(2\pi d)) + \Delta (p^6/(2\pi)) (d_c/d)^6}. 
\label{rhoDints}
\end{align}

Where $\Delta = D_\eta D_\sigma/(D_\eta+D_\sigma)^2$ is independent of $d$ and $T$ for the Fermi surface coupled to $U(1)$ gauge field. We can neglect the small $d_e$. At large $d\gg d_c$, we consider the first term in $\rho_D^\sigma$: The exponential in the numerator implies that at large $d\gg d_c$, the region of interest is $p\lesssim 1$. In this region, the first term of the denominator dominates the second. We can thus approximate the integrand by $p^3/(2\pi)$. The integral is cut off when the two terms in the denominator become comparable, i.e. when 
\begin{equation}
p \approx \frac{3}{2}W\left(\frac{2}{3}\left(\frac{2\pi Q^{(0)2}d^3}{M(D_\eta+D_\sigma)^2}\right)^{1/3}\right) \sim \ln \left(\frac{d}{d_c}\right),
\end{equation}
where $W(x)$ is the Lambert $W$ function, which has the property $W(x\gg1) \approx \ln x - \ln \ln x$. The same strategy can be used to do the integral for $\rho_D^\eta$. The second term in $\rho_D^\sigma$ falls off much faster faster with $d$ than the first term. It is also non-singular as $T\rightarrow0$. We can thus ignore it. We then obtain
\begin{align}
&\rho_D^\sigma \sim \frac{k_BT\sigma^{Q(0)}M}{Q^{(0)4}d^4}\ln^4\left(\frac{d}{d_c}\right),  \\
&\rho_D^\eta \sim \frac{k_BT(\eta^{(0)}+\zeta^{(0)})\varepsilon}{Q^{(0)4}d^5}\ln^5\left(\frac{d}{d_c}\right),
\label{rhodhyd}
\end{align}
which translates to (\ref{rhoDfinal}) of the main text. At small separations $d\ll d_c$ we can set the exponential factors in (\ref{rhoDints}) to unity, without making the integrals UV divergent. Then, all the contributions become independent of $d$, and after considering the $T$ dependence of all quantities, we obtain $\rho_D^\sigma \sim T^{19/9}$ and $\rho_D^\eta \sim T^{23/9}$ for the Fermi surface coupled to $U(1)$ gauge field with dynamical critical exponent $z=3$.

For this Fermi surface coupled to $U(1)$ gauge field, we also have~\cite{Eberlein2016, Nave2007}
\begin{equation}
\alpha^{Q(0)} \sim T^{-2/3},~~~\bar{\kappa}^{Q(0)} \sim T^{1/3}.
\end{equation}
Thus, the correlators of thermal and thermoelectric noise in (\ref{noisecorrel}) of the main text are smaller than the correlators of charge and viscous noise at low temperatures, and may be ignored. As far as corrections from thermal current flow to our results are concerned, we observe that from~(\ref{rhoDintsU}) the diffusion constant $D_\sigma$ associated with the microscopic conductivity $\sigma^{Q(0)}$ always multiplies additional powers of $k$ in the integrands, and thus only provides subleading contributions at large values of $d$. This is also true for contributions of diffusion constants associated with the microscopic thermal and thermoelectric conductivities $\bar{\kappa}^{Q(0)}$ and $\alpha^{Q(0)}$. Thus, at large separations, thermal currents influence the charge current flow in the dragged layer only through the entropy density, given by $S^{(0)}\sim T^{2/3}$~\cite{Eberlein2016,Halperin1993}, which, by virtue of vanishing at low temperatures, can only provide corrections to our results that scale as subleading powers of $T$. All this may be verified by a very lengthy brute-force computation, which we choose not to include as it is not important for our main results.

For a general dynamical critical exponent $z$, the arguments of~\cite{Eberlein2016,Eberlein2017} imply that 
\begin{equation}
\sigma^{Q(0)} \propto \left(\frac{E_F}{k_BT}\right)^{2/z}\frac{e^2}{\hbar},~~~\eta^{(0)} \sim \zeta^{(0)} \propto \left(\frac{E_F}{k_BT}\right)^{2/z}\hbar k_F^2.
\label{fscrz}
\end{equation}
We then get that $\rho_D\sim \rho_D^\sigma \sim T^{1-2/z}\ln^4 T$ for $d\gg d_c$, and $\rho_D \sim \rho_D^\sigma \sim T^{1+10/(3z)}$ for $d\ll d_c$. Thus the curvature of the $\rho_D$ vs $T$ plot (Fig.~\ref{rhoDscaling} of the main text) still changes from positive to negative as $d$ is increased, and this change is more pronounced for $z$ closer to 2 than to 3. In the HLR theory, $z$ ranges between 2 (For long-range Coulomb interactions of composite fermions) and 3 (For short-range interactions of composite fermions). The scaling of $\rho_D$ with $T$ in the hydrodynamic regime at large $d$ can thus possibly yield some insight into the type of composite-fermion interactions leading to non-Fermi liquid behavior.

\section{Remarks on Coulomb drag in the $\nu=1/2$ quantum Hall state}

This section serves to provide a brief summery of the main results from \cite{Ussishkin:1997}. Consider a double-layer system of quantum Hall states at half-filling. We want to study Coulomb drag in this setting by adopting the theoretical framework of composite fermions (CF) based on the seminal work of HLR. We assume that both layers are very clean so that the intralayer mean free path $l_{cf}$ of the composite fermions is limited only by the scattering off the gauge field. For the case when $l_{cf}$ is large, drag was studied by Ussishkin and Stern \cite{Ussishkin:1997}. Large $l_{cf}$ implies ``collisionless" transport with respect to intralayer collisions. In this case drag resistivity can be described by the conventional expression 
\begin{equation}
\rho_D=\frac{\rho_Q}{8\pi^2Tn^2}\sum_q\int^{\infty}_{0}\frac{d\omega}{\sinh^2(\omega/2T)}q^2|W_{q,\omega}|^2(\mathrm{Im}\Pi_{q,\omega})^2.
\end{equation}    
The screened interlayer potential can be written in the form 
\begin{equation}
W=\frac{(V+U)/2}{1+\Pi(V+U)}-\frac{(V-U)/2}{1+\Pi(V-U)},
\end{equation}
where $V=2\pi e^2/\varepsilon q$ and $U=Ve^{-qd}$. The polarization operator $\Pi_{q,\omega}$ entering the drag formula  should be computed from the CF picture by the following prescription. According to HLR theory one starts out from the matrix equation  
\begin{equation}
(\hat{\Pi}^e)^{-1}=\hat{C}+(\hat{\Pi}^{cf})^{-1},
\end{equation}
where $\hat{\Pi}^e$ is the electronic single layer response function, whereas $\hat{\Pi}^{cf}$ is the polarization (density-density and current-current) of composite fermions, and $\hat{C}$ is the matrix of attached flux. Explicitly we have
\begin{equation}
\hat{C}=\left(\begin{array}{cc}0 & 2\pi i\phi/q \\ -2\pi i\phi/q & 0\end{array}\right), \quad 
\hat{\Pi}^{cf}=\left(\begin{array}{cc}\Pi^{cf}_{00} & 0 \\ 0 & \Pi^{cf}_{11}\end{array}\right).
\end{equation} 
In the limit of $q/k_F\ll1$ and also $\omega\ll v_Fq$, the random-phase-approximation results for density and current responses (including Chern-Simons contributions) of composite fermions are 
\begin{equation}
\Pi^{cf}_{00}\approx\frac{m}{2\pi}, \qquad \Pi^{cf}_{11}\approx-\frac{q^2}{24\pi m}+\frac{i\omega k_F}{2\pi q}.
\end{equation}
The finite value of $\Pi^{cf}_{00}$ reflects the compressibility of the system, whereas $\Pi^{cf}_{11}$ reflects Landau diamagnetism (the real part) and Landau damping (the imaginary part). Inverting $\Pi^e$ and taking its density component gives
\begin{equation}
\Pi_{q,\omega}=\Pi^{e}_{00}=\frac{\Pi^{cf}_{00}}{1-\Pi^{cf}_{00}\Pi^{cf}_{11}(2\pi\phi/q)^2}=\frac{\chi q^3}{q^3-2\pi i\phi^2\chi\omega k_F}.
\end{equation} 
It will be convenient for us to introduce the dimensionless variables $x=qd$ and $y=\omega/T$ such that 
\begin{equation}
\Pi_{x,y}=\frac{\chi x^3}{x^3-i\alpha_Ty},\quad \alpha_T=2\pi\phi^2\chi Tk_Fd^3.
\end{equation}
If we introduce an inverse Thomas-Fermi screening radius of composite fermions as $\kappa=2\pi e^2\chi/\varepsilon$, then $\alpha_T$ can be rewritten as $\alpha_T=\phi^2(\kappa d)(k_Fd)T/(e^2/\varepsilon d)$. To proceed with the computation of $\rho_D$ we notice that 
\begin{equation}
\mathrm{Im}(\Pi) |W|=-\mathrm{Im}(\Pi^{-1}) \left|\frac{U}{(\Pi^{-1}+V+U)(\Pi^{-1}+V-U)}\right|.
\end{equation} 
In terms of dimensionless variables 
\begin{eqnarray}
&&U=\frac{\kappa d}{\chi}\frac{e^{-x}}{x},\quad \mathrm{Im}\Pi^{-1}=-\frac{\alpha_Ty}{\chi x^3},\quad \\
&&\frac{U}{(\Pi^{-1}+V+U)(\Pi^{-1}+V-U)}=\frac{\chi\kappa dx^5e^{-x}}{[x^3-i\alpha_Ty+\kappa dx^2(1+e^{-x})]
[x^3-i\alpha_Ty+\kappa dx^2(1-e^{-x})]}.
\end{eqnarray}
Now setting up an integral for the drag resistivity in dimensionless notation we obtain the following expression 
\begin{equation}
\frac{\rho_D}{\rho_Q}=\frac{\alpha^2_T}{32\pi^4}\frac{(\kappa d)^2}{(nd^2)^2}\int^{\infty}_{0}\frac{y^2dy}{\sinh^2(y/2)}\int^{\infty}_{0}\frac{x^7e^{-2x}dx}{|x^3-i\alpha_Ty+\kappa dx^2(1+e^{-x})|^2|x^3-i\alpha_Ty+\kappa dx^2(1-e^{-x})|^2}.
\end{equation}
At the lowest temperatures $T\to0$ we have $\alpha_T\ll1$. Coulomb screening and thermal factors set the typical scales of momentum and energy $x\sim y\sim 1$, however the pole structure of the denominator is dominated by the momentum range $x\sim\sqrt[3]{\alpha_T}\ll1$. Because of that we can make the following approximations: $x^3-i\alpha_Ty+\kappa dx^2(1+e^x)\approx 2\kappa dx^2$,  $x^3-i\alpha_Ty+\kappa dx^2(1-e^x)\approx (1+\kappa d)x^3-i\alpha_Ty$ and $e^{-2x}\approx 1$. As a result, the previous expression simplifies to 
\begin{equation}
\frac{\rho_D}{\rho_Q}=\frac{\alpha^2_T}{128\pi^4}\frac{1}{(nd^2)^2}\int^{\infty}_{0}\frac{y^2dy}{\sinh^2(y/2)}\int^{\infty}_{0}\frac{x^3dx}{|(1+\kappa d)x^3-i\alpha_Ty|^2}.
\end{equation}
Upon rescaling of $x$ these integrals can be brought to the form 
\begin{equation}
\frac{\rho_D}{\rho_Q}=\frac{1}{128\pi^4}\frac{1}{(nd^2)^2}\left(\frac{\alpha_T}{1+\kappa d}\right)^{4/3}\int^{\infty}_{0}\frac{y^{4/3}dy}{\sinh^2(y/2)}\int^{\infty}_{0}\frac{x^3dx}{x^6+1},
\end{equation} 
where the $x$-integral is equal to $\pi/(3\sqrt{3})$ while the $y$-integral is equal to $4\Gamma(7/3)\zeta(4/3)$. 
Combining all factors and using $k_F=\sqrt{4\pi n}$ we find 
\begin{equation}
\frac{\rho_D}{\rho_Q}=\frac{\Gamma(7/3)\zeta(4/3)}{6\pi\sqrt{3}}\left(\frac{T}{T_0}\right)^{4/3},\quad T_0=\frac{e^2}{\phi^2\varepsilon d}(k_Fd)^2\left(1+\frac{1}{\kappa d}\right). 
\label{rhodus} 
\end{equation}

The length scale for the Coulomb potential induced on one layer by density fluctuations in the other layer is roughly given by the interlayer separation $d$. For our hydrodynamic analysis to be applicable, $d$ must be much larger than the intralayer mean free path $l_{cf}$. Thus, at small values of $d$,~(\ref{rhodus}) is more likely to be applicable. Since $\rho_D$ falls off with $d$ only as $d^{4/3}$ in,~(\ref{rhodus}), but as $d^4$ in the hydrodynamic result~(\ref{rhodhyd}), the measured spacing dependence of $\rho_D$ can also possibly be used to deduce the pertinent transport regime. Note that the $T$ dependence of $\rho_D$ is $T^{4/3>1}$ in~(\ref{rhodus}), so the curvature of the $\rho_D$ vs $T$ plot should still switch from positive to negative as $d$ is increased even if there is a crossover from the collisionless to the hydrodynamic regime. 

\end{document}